\documentclass[11pt]{article}

\PassOptionsToPackage{table,xcdraw}{xcolor}
\usepackage[final]{acl}  

\usepackage{times}
\usepackage{latexsym}

\usepackage[T1]{fontenc}

\usepackage[utf8]{inputenc}

\usepackage{microtype}

\usepackage{inconsolata}

\usepackage{graphicx}
\usepackage{arydshln}

\newcommand\blfootnote[1]{%
  \begingroup
  \renewcommand\thefootnote{}\footnote{#1}%
  \addtocounter{footnote}{-1}%
  \vspace{-4.25mm}%
  \endgroup
}

\usepackage{amsmath,amssymb,amsfonts}
\usepackage{algorithm}
\usepackage{algorithmicx}
\usepackage{algpseudocode}
\usepackage{booktabs}
\usepackage{multirow}
\usepackage{enumitem}

\title{MeanAudio: Fast and Faithful Text-to-Audio Generation with Mean Flows}
\author{
\vspace{0.2cm}
        \textbf{Xiquan Li\textsuperscript{\rm 1,2$\dag$}},
        \textbf{Junxi Liu\textsuperscript{\rm 1$\dag$}}, 
        \textbf{Yuzhe Liang\textsuperscript{\rm 1}},
        \textbf{Zhikang Niu\textsuperscript{\rm 1}}, 
        \textbf{Wenxi Chen\textsuperscript{\rm 1}},
        \textbf{Xie Chen\textsuperscript{\rm 1}} \\
         \textsuperscript{\rm1}MoE Key Lab of Artificial Intelligence, X-LANCE Lab \\ Department of Computer Science, Shanghai Jiao Tong University, China \\
\textsuperscript{\rm2}SJTU Paris Elite Institute of Technology, Shanghai Jiao Tong University, China
\\ \texttt{\{mtxiaoxi55, chenxie95\}@sjtu.edu.cn}
}

\begin{document}
\maketitle

\begin{abstract}
Recent years have witnessed remarkable progress in Text-to-Audio Generation (TTA), providing sound creators with powerful tools to transform inspirations into vivid audio.
Yet despite these advances, current TTA systems often suffer from slow inference speed, which greatly hinders the efficiency and smoothness of audio creation.
In this paper, we present MeanAudio, a fast and faithful text-to-audio generator capable of rendering realistic sound with only one function evaluation (1-NFE). 
MeanAudio leverages: (i) the MeanFlow objective with guided velocity target that significantly accelerates inference speed, (ii) an enhanced Flux-style transformer with dual text encoders for better semantic alignment and synthesis quality, and (iii) an efficient instantaneous-to-mean curriculum that speeds up convergence and enables training on consumer-grade GPUs.
Through a comprehensive evaluation study, we demonstrate that MeanAudio achieves state-of-the-art performance in single-step audio generation. 
Specifically, it achieves a real-time factor (RTF) of 0.013 on a single NVIDIA RTX 3090, yielding a 100x speedup over SOTA diffusion-based TTA systems. 
Moreover, MeanAudio also shows strong performance in multi-step generation, enabling smooth transitions across successive synthesis steps. 
\end{abstract}

\blfootnote{$^\dag$Equal contributions.}
\blfootnote{$^\ddag$Code: \url{https://github.com/xiquan-li/MeanAudio} Demo: \url{https://MeanAudio.github.io/}}


\section{Introduction}
Text-to-Audio Generation (TTA) \cite{liu2023audioldm, ghosal2023text, huang2023make} aims to synthesize diverse auditory content from textual prompts. 
By translating language into sound, TTA models unlock a broad spectrum of real-world applications, including virtual reality, gaming, film post-production, and human-computer interaction.

In pursuit of stronger audio generators, recent studies have advanced along three complementary directions: architectural innovations \cite{huang2023make2, hai2024ezaudio, hung2024tangoflux, haji2024taming, evans2025stable, lee2024etta, valle2025fugatto}, scaling data and model parameters \cite{kong2024improving, yuan2024sound, liu2024audioldm2, haji2024taming}, and employing novel training objectives \cite{majumder2024tango2, liao2024baton}.
These approaches have substantially improved generation quality and controllability, as evidenced by the Inception Score (IS) on AudioCaps \cite{kim2019audiocaps} rising from 8.17 \cite{liu2023audioldm} to 12.81 \cite{hung2024tangoflux}.
However, despite these improvements, current TTA models often suffer from slow inference speed, where they typically require seconds to minutes to synthesize a single audio sample. 
Such latency, stemming from the iterative sampling process in flow and diffusion-based models, not only impedes deployment in time-sensitive scenarios such as virtual assistants and interactive gaming, but also hinders the creative workflow of sound creators.

\begin{figure}
    \centering
    \includegraphics[width=1\linewidth]{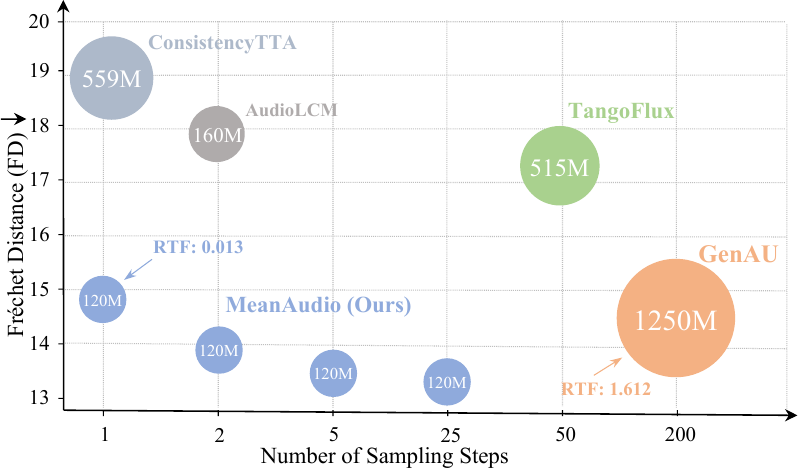}
    \caption{MeanAudio achieves state-of-the-art single-step generation performance with a real-time factor (RTF) of 0.013, offering a 100x speedup over existing diffusion-based TTA systems. It also demonstrates strong performance in multi-step generation, despite using only 120M parameters. }
    \label{fig:performance}
    \vspace{-0.6cm}
\end{figure}

To accelerate the inference speed of TTA models, recent studies \cite{liu2024audiolcm, liu2024flashaudio, saito2025soundctm, bai2023consistencytta} have primarily focused on diffusion distillation \cite{song2023consistency}.
In this paradigm, the number of diffusion sampling steps is reduced by distilling a pre-trained teacher model into a few-step student generator. 
As such, the student generator learns to synthesize audio in a few steps by modeling the flow trajectory of their multi-step teachers. 
While these models have achieved promising performance in single-step and few-step audio generation, they are inherently limited by the rigid consistency constraints and their reliance on teacher models. 
Moreover, the distillation-based approaches are often computationally expensive, as online methods require holding 2-3 full models in memory simultaneously, and offline methods rely on large-scale generation and storage of teacher trajectories before training.

In this paper, we present MeanAudio, a MeanFlow-based \cite{geng2025mean} fast and faithful text-to-audio generator capable of rendering high-quality audio with only 1 iteration. 
To improve synthesis quality, MeanAudio leverages an enhanced Flux-Style \cite{flux2024} flow transformer with dual text encoders, facilitating realistic and instruction-adherent audio generation. 
To accelerate inference, MeanAudio regresses the average velocity field during training, enabling direct mapping from the start to the endpoint of the flow trajectory. 
By further integrating classifier-free guidance (CFG) \cite{ho2022classifier} into the training objective, it achieves guided sampling without additional computational cost.
Moreover, we introduce an instantaneous-to-mean learning curriculum with flow-field mix-up, which anchors the model in the instantaneous velocity field before progressively adapting to the average velocity field.
This strategy proves beneficial for improving both training efficiency and generation performance across single- and multi-step inference.


Through extensive experiments, we show that MeanAudio achieves state-of-the-art (SOTA) performance in single-step TTA generation. 
Notably, it achieves a real-time factor (RTF) of 0.013, corresponding to 100x speedup over the best open-sourced diffusion-based TTA system, GenAU \cite{haji2024taming}, which requires 200 sampling steps.
Beyond single-step generation, MeanAudio also demonstrates competitive performance in multi-step synthesis, all within a compact 120M-parameter model that can be efficiently trained in three days on four NVIDIA RTX 3090 GPUs.
In addition, to uncover best practices for building MeanFlow-accelerated audio generators, we conduct a comprehensive ablation study that highlights the importance of architectural choices, training strategies, and flow configurations.
To summarize, our main contributions are as follows: 



\begin{itemize}[itemsep=2pt, topsep=2pt, parsep=0pt, partopsep=0pt]
   \item We present \textbf{MeanAudio}, the first text-to-audio generator that learns MeanFlows for fast and faithful sound synthesis. 
    \item We design an improved Flux-Style flow transformer with dual text encoders, enabling high-quality and prompt-adherent audio generation.
    \item We introduce an instantaneous-to-mean curriculum with flow mix-up that facilitates stable training and rapid convergence. 
    \item Extensive experiments show that MeanAudio achieves SOTA results in single-step generation and competitive performance in multi-step generation.
    \item We provide comprehensive experimental analysis that highlights best practices for building MeanFlow-based audio generators, paving the way toward faster and stronger audio models.
\end{itemize}
We will fully release the MeanAudio codebase and model weights to facilitate future research on efficient and high-quality text-to-audio generation.

\begin{figure*}[t]
    \centering
    \vspace{-0.3cm}
    \includegraphics[width=\linewidth]{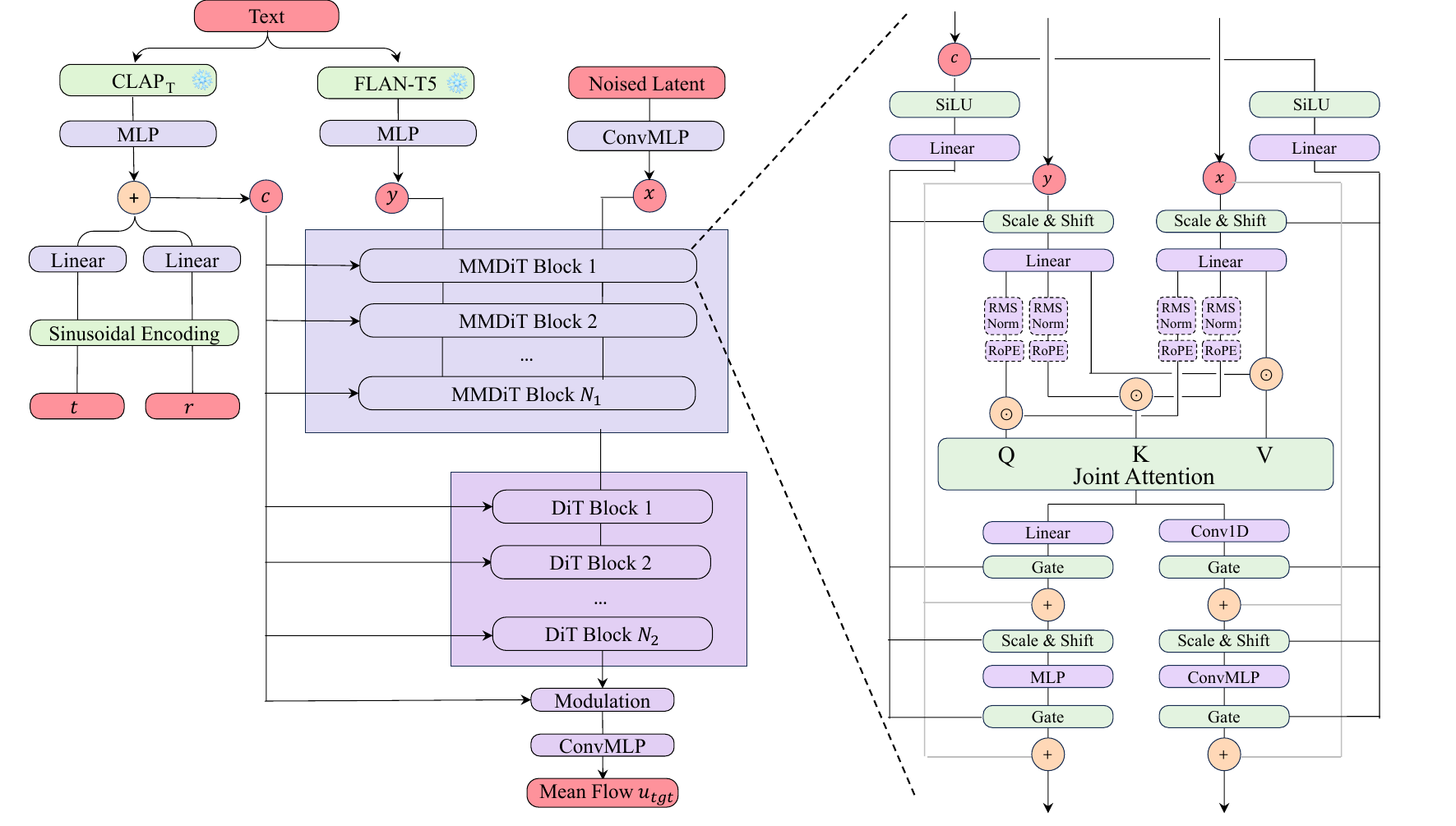}
    \vspace{-0.8cm}
    \caption{Model architecture overview: MeanAudio combines $N_1$ multi-modal (MMDiT) blocks and $N_2$ single-modal (DiT) blocks to construct the flow transformer. 
    It leverages the joint attention in multi-modal blocks to integrate FLAN-T5’s fine-grained text embeddings, and employs AdaLN to inject CLAP’s global conditioning. 
    \vspace{-0.3cm}
    }
    \label{fig:mean_audio}
\end{figure*}

\section{Preliminaries}

\subsection{Conditional Flow Matching}
Flow Matching \cite{liu2022flow, lipman2022flow, albergo2022building} is a powerful generative model that learns to match the flows between two probabilistic distributions. 
Given data $x \sim p_\text{data}(x)$, prior $\epsilon \sim p_\text{prior}(\epsilon)$, the optimal transport flow path can be constructed as: $x_t = (1-t)x + t \epsilon$, and the conditional velocity is thus given by $v_t = \frac{dx_t}{dt} = \epsilon - x$.  
At training time, the objective is to find a neural network $f_\theta$ which minimizes the conditional flow matching loss \cite{lipman2022flow}: 
\begin{equation}
\label{eq:fm_loss}
    \mathcal{L}_\text{CFM} = \mathbb{E}_{t, x, \epsilon}\|f_\theta (t, x_t) - v_t\|^2
\end{equation}

During sampling, we randomly draw noise $\epsilon$ from $p_\text{prior}$ and solve the ordinary differential equation (ODE) defined below:
$$
dx_t = -f_\theta (t, x_t)dt
$$
The solution is thus given by $x_r = x_t - \int_r^tf_\theta(\tau, x_\tau)d\tau$, where $r$ denotes another time step. 
During implementation, we can use numerical methods (e.g. Euler Method) to approximate this integration. 

\subsection{Mean Flows for Generative Modeling}
To accelerate the inference speed of Flow Matching, Mean Flows \cite{geng2025mean} proposed to regress the average velocity field during training, allowing high-quality single-step generation. 
Specifically, given a time interval $[r, t]$, the average velocity within it is defined as:
$u(x_t, r, t) \triangleq \frac{1}{t-r}\int_r^t v(x_\tau, \tau) d\tau$. 
By differentiating both sides with respect to $t$ and re-arranging terms, we obtain the \textit{Mean Flow Identity}, which describes the relation between $v$ and $u$: 
\begin{equation*}
u(x_t, r, t) = v_t - (t-r)\frac{d}{dt}u(x_t, r, t)
\end{equation*}
We then encourage $f_\theta$ to satisfy this identity by minimizing the mean flow objective: 
\begin{equation}
\label{eq:mf_loss}
    \mathcal{L}_\text{MF}= \mathbb{E}_{t, r, x, \epsilon} \| f_\theta(x_t, r, t) - \text{sg}(u_\text{tgt})\|^2 
\end{equation}
Where $u_\text{tgt}=v_t  - (t-r)\frac{d}{dt}f_\theta(x_t, r, t)$, and $\text{sg}(\cdot)$ denotes the stop-gradient operation.  
Note that this total derivative can be expanded by its partial components, corresponding to a Jacobian-Vector Product (JVP): 
$\frac{d}{dt}f_\theta(x_t, r, t) = v_t\partial_x f_\theta+ \partial_t f_\theta$. 
When $r=t$, the mean flow objective becomes the vanilla flow matching objective. 

During sampling, the time integral in CFM can be replaced by the average velocity, leading to: 
$$
x_r = x_t - (t-r)f_\theta(x_t, r, t)
$$
In particular, in single-step generation, we have: 
$
x_0 = x_1 - f_\theta(x_1, 0, 1)
$
where $x_1 = \epsilon \sim p_\text{prior}(\epsilon)$. 

\section{MeanAudio}

As illustrated in Figure \ref{fig:mean_audio}, MeanAudio employs an enhanced Flux-Style flow transformer to learn average velocity in the latent space conditioned on the textual prompt and timestep embeddings.
In this section, we first describe the architectural design of MeanAudio, and then introduce our strategies for accelerating training and inference.

\subsection{Audio Encoding}
Following prior works \cite{liu2023audioldm, ghosal2023text}, we model the generative process in the latent space to improve computational efficiency. Specifically, we apply the short-time Fourier transform (STFT) to audio waveforms and extract the magnitude component as mel spectrograms \cite{stevens1937scale}. These spectrograms are then encoded into latent representations $x$ using a pre-trained variational autoencoder (VAE) \cite{kingma2013auto}. During inference, the generated latents are decoded back into spectrograms via the VAE and subsequently converted to audio waveforms using a pre-trained vocoder \cite{lee2022bigvgan}. We employ a 1D convolution-based VAE for its superior capacity to model frequency- and length-variable representations. For a 10-second audio input, the autoencoder produces a latent sequence of 312 tokens, each with a hidden dimension of 20.

\subsection{Enhanced Flow Transformer}
We design an enhanced Flux-Style flow transformer to improve MeanAudio's synthesis quality and prompt adherence. 
Specifically, we combine $N_1$ multi-modal transformer blocks (MMDiT) \cite{esser2024scaling} with audio/text branches and $N_2$ audio-only DiT \cite{peebles2023scalable} blocks to construct our flow transformer.
To further boost generation quality, we employ several refinements: Firstly, we use ConvMLP rather than vanilla MLPs in the audio stream of MeanAudio. ConvMLP uses 1D convolutions (kernel size = 3 and padding = 1) rather than linear layers, demonstrating stronger performance in capturing local temporal structure \cite{cheng2024taming}.
Secondly, we apply rotary positional embedding (RoPE) \cite{su2024roformer} on the queries and keys in both the audio and text branches. 
Unlike absolute position embeddings, RoPE models the relative distances and is beneficial for variable-length audio generation. 
Thirdly, we use RMSNorm \cite{zhang2019root} with learnable scales in attention calculation to enable stable and efficient training.
Lastly, we adopt the SwiGLU \cite{shazeer2020glu} activations instead of ReLU \cite{agarap2018deep} in the MLP layers. 

\subsection{Model Conditioning}
MeanAudio is conditioned on a textual prompt and time steps to render faithful audio signals. 
For textual conditioning, we leverage FLAN-T5 \cite{chung2024scaling} and CLAP \cite{wu2023large} to extract caption embeddings. 
FLAN-T5 is an instruction-tuned large-language model (LLM) capable of producing fine-grained token embeddings. 
Meanwhile, CLAP is pre-trained on large-scale audio-text dataset and can offer global acoustic-aligned text embeddings. 
Denote $y_\text{T5} \in \mathbb{R}^{N\times d_\text{T5}}$ as the embedding extracted by FLAN-T5, where $N$ and $d_\text{T5}$ represent the number of tokens and model's output dimension. 
We feed $y_\text{T5}$ into the text branch of MMDiT, where multi-modal joint attention learns cross-modal alignment between text and audio.
Furthermore, let $y_\text{CLAP} \in \mathbb{R}^{1\times d_\text{CLAP}}$ represent the global text embedding obtained from the CLAP text encoder (CLAP$_\text{T}$). We project this embedding through an MLP and combine it with the extracted timestep features to form the global condition $c = t_\text{emb} + r_\text{emb} + y'_\text{CLAP}$. 
This global information is then injected into the model via the scales and biases of adaptive layer normalization (AdaLN) layers.
While FLAN-T5 can capture fine-grained textual details at the token level, CLAP contributes holistic, audio-grounded semantic information. Together, they provide rich and balanced conditions that improve both the fidelity and semantic alignment of the generated audio.



\subsection{Integrated Classifier-Free Guidance}
Classifier-free guidance (CFG) \cite{ho2022classifier} is a widely adopted technique for achieving controllable generation.
However, using CFG during sampling doubles the number of function evaluations (NFE), as both class-conditional and unconditional model outputs should be computed.
To eliminate the additional cost associated with guided sampling, MeanAudio integrates CFG into the training target. 
Specifically, define $v_t^\text{cfg}$ as the estimated instantaneous velocity field with guidance, which can be expressed as follows: 
\begin{align}
v_t^\text{cfg} 
= \omega v_t 
&+ \underbrace{\kappa f_\theta(x_t, t, t | \textbf{C})}_{\text{cls. conditional}} \notag \\
&+ \underbrace{(1-\omega-\kappa)f_\theta(x_t, t, t|\varnothing)}_{\text{cls. unconditional}}
\end{align}
Here, $\kappa$ is a mixing factor which combines both class-conditional and unconditional predictions into the guided field, resulting in an effective guidance scale of $\omega' = \frac{\omega}{1-\kappa}$. 
Similarly, we also expose the trainable network $f_\theta$ with class-unconditional inputs, where we randomly drop $\textbf{C}$ with 10\% probability, following \cite{ho2022classifier}. 
By replacing $v_t$ with $v_t^\text{cfg}$ in $u_\text{tgt}$, we obtain the average velocity target with guidance, which can be formulated as: 
$$
u_\text{tgt}^\text{cfg} = v^\text{cfg}_t -  (t-r)\frac{d}{dt}f_\theta(x_t, r, t)
$$
By regressing $u_\text{tgt}^{\text{cfg}}$, MeanAudio directly learns the guidance during training, thus avoiding the need for an additional forward pass during generation.
The training and inference procedure of MeanAudio are illustrated in Algorithm \ref{alg:meanaudio_training} and \ref{alg:meanaudio_inference}.

\begin{algorithm}[t]
\caption{\textbf{MeanAudio Training}}
\label{alg:meanaudio_training}
\begin{algorithmic}[1]
\State \textbf{Input:} Encoded audio latent: $x$, textual conditions $\textbf{C}$, flow transformer $f_\theta$. 
\State Sample $t, r \sim \text{lognorm}(\mu, \sigma)$, $\epsilon \sim \mathcal{N}(0, I)$

\State $x_t \gets (1 - t) \cdot x + t \cdot \epsilon$
\State $v_t \gets \epsilon - x$
\State Compute model output $f_\theta(x_t, r, t)$ and its derivative $\frac{df_\theta}{dt}$ via JVP: \\
$f_\theta(x_t, r, t), \frac{df_\theta}{dt}\gets \text{JVP}(f_\theta, (x_t, r, t), (v, 0, 1))$
\State Compute the guided instantaneous velocity estimate:
\State \hskip1em $v_t^\text{cfg} \gets \omega \cdot v_t + \kappa \cdot f_\theta(x_t, t, t, \varnothing) + (1 - w - \kappa) \cdot f_\theta(x_t, t, t, \textbf{C})$
\State Compute the guided average velocity estimate:
\State \hskip1em $u_{\text{tgt}}^\text{cfg} \gets  v_t^\text{cfg} - (t - r) \cdot \frac{d}{dt}f_\theta$
\State $\text{loss} \gets \|f_\theta(x_t, r, t) - \text{sg}(u_{\text{tgt}})\| ^2$
\State \textbf{Output:} loss
\end{algorithmic}
\end{algorithm}

\begin{algorithm}[t]
\caption{\textbf{MeanAudio Inference}}
\label{alg:meanaudio_inference}
\begin{algorithmic}[1]
\State \textbf{Input:} Trained network $f_\theta$, textual conditions $\textbf{C}$
\State Sample noise: $\epsilon \sim \mathcal{N}(0, I)$
\State $x \gets \epsilon$

\For{$i = 0$ to $N-1$}
    \State $x = x - (t_{i+1} - t_{i}) \cdot f_\theta(x, t_i, t_{i+1}, \textbf{C})$
\EndFor

\State \textbf{Output:} $x$
\end{algorithmic}
\end{algorithm}

\subsection{Stabilizing Flow Fields}
\label{sec:training_strategy}
Although the \textit{Mean Flow Identity} provides an effective training target for learning fast single-step generation, we found that directly modeling audio latent with Eq. \ref{eq:mf_loss} results in unstable training, slow convergence, and poor multi-step generation.
This may be due to several factors:
Firstly, the MeanFlow objective defined in Eq.~\ref{eq:mf_loss} focuses solely on learning the average velocity, which may cause the model to neglect the underlying instantaneous field that serves as the foundation for Mean Flow \cite{peng2025flow}.
Secondly, the training target $u^\text{cfg}_\text{tgt}$ is defined by the model’s own derivative. However, a randomly initialized model may fail to provide effective guidance at the beginning, resulting in slow convergence.

\begin{figure}
    \centering
    \includegraphics[width=1\linewidth]{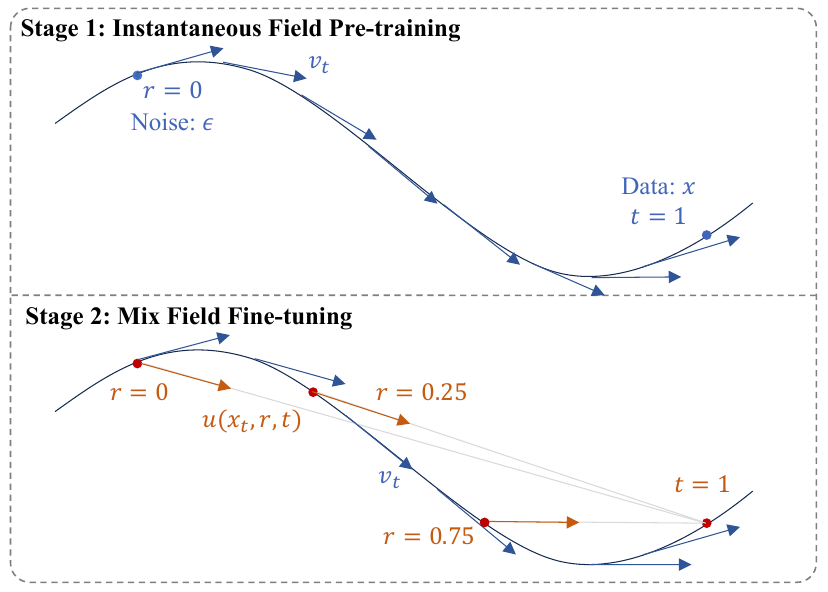}
    \caption{
    Illustration of the underlying instantaneous field $v_t$, which models tangents of the flow trajectory, and the average velocity field $u(x_t, r, t)$, which captures long displacements. 
    Our training curriculum encourages the model to first learn the foundational instantaneous dynamics, then gradually adapt to mean flows for fast and faithful generation.
    }
    \label{fig:flow_trajctory}
    \vspace{-0.5cm}
\end{figure}

To address this issue, we propose an instantaneous-to-mean curriculum with flow field mixup to improve training stability and efficiency. 
As illustrated in Figure \ref{fig:flow_trajctory}, our curriculum comprises two stages: In the first stage, the model is trained on large-scale, weakly-labeled audio-text datasets to learn the instantaneous velocity field, whose loss is defined in Equation \ref{eq:fm_loss}.
This stage establishes a strong initialization, allowing the model to first capture the underlying foundational dynamics. 
In the second stage, the model is fine-tuned on a smaller, high-quality dataset to learn the mean velocity. 
In this stage, we adopt the strategy from \cite{geng2025mean}, where we blend the instantaneous and average fields by randomly setting $r = t$. As illustrated in Eq.~\ref{eq:mf_loss}, this operation degenerates the MF objective into standard flow matching.
As such, the network can provide an effective derivative approximation during fine-tuning by leveraging the knowledge acquired in pre-training.
Furthermore, by combining two flows, the model can stably adapt to the average field for fast few-step generation, while preserving the multi-step synthesis performance through adherence to instantaneous velocity.
We will demonstrate the effectiveness of our training curriculum in the experimental section. 

\begin{table*}[t]
\centering
\resizebox{1\linewidth}{!}{
\begin{tabular}{lcccccccc}
\toprule
\textbf{Model} & \textbf{Prams}$^\ast$ & \textbf{NFE} & \textbf{FAD} ↓ & \textbf{FD} ↓ &  \textbf{KL} ↓ & \textbf{IS} ↑ & \textbf{CLAP}  ↑ & \textbf{RTF}$^\dag$ ↓ \\
\midrule
\textbf{Diffusion- and Flow-based TTA Models} 
\\
AudioLDM-L-Full \cite{liu2023audioldm}$^\ddag$            & 739M  & 200   & 4.32 & 29.50  & 1.68 & 8.17   & 0.208 &  2.935\\ 
Tango-Full-FT \cite{ghosal2023text}$^\ddag$          & 866M  & 200    & 2.68 & 15.64  & 1.24 & 8.78 & 0.291 & 2.382\\
EzAudio-XL \cite{hai2024ezaudio}$^\ddag$               & 875M  & 200   & 3.64 & 14.98  & 1.29 & 11.38 & 0.314 & 1.718\\
Stable Audio Open \cite{evans2025stable}$^\ddag$ & 1050M & 200 & 4.19 & 39.14 & 2.36 & 10.07 & 0.209 & 2.867 \\
TangoFlux \cite{hung2024tangoflux}$^\ddag$         & 516M  & 50   & 2.41 & 20.65 & 1.27 & \underline{12.81} & \underline{0.318} & \underline{0.449} \\
IMPACT-Large \cite{huang2025impact}$^{**}$ & 427M & 64 & \textbf{1.17} & 14.72 & \textbf{1.07} & 10.53 & - & - \\
ETTA \cite{lee2024etta}$^{**}$ &  1440M & 100 & - & \textbf{13.12} & 1.42 & \textbf{14.36} & - & - \\
GenAU-Large-Full \cite{haji2024taming}$^\ddag$      & 1250M  & 200 & \underline{2.07} &  14.58 & 1.36 & 10.43 & 0.300 & 1.612 \\
\midrule
\textbf{Accelerated TTA Models} \\
AudioLCM \cite{liu2024audiolcm}$^\ddag$  & 160M & \textit{1} & \textit{4.70} & \textit{24.59} & \textit{1.67} & \textit{8.04} & \textit{0.200} & \textit{0.023} \\
& & 2 & 2.16 & 19.81 & 1.46 & 10.05 & 0.245 & 0.026 \\
& & 5 & 1.76 & 20.12 & 1.46 & 10.36 & 0.256 & 0.029\\
ConsistencyTTA \cite{bai2023consistencytta}$^{\diamondsuit, \ddag}$  & 559M & \textit{1} & \textit{\underline{2.31}} & \underline{\textit{22.16}} & \underline{\textit{1.44}} & \underline{\textit{9.15}} & \underline{\textit{0.268}} & \underline{\textit{0.017}} \\
& & 2 & 2.68 & 22.25 & 1.42 & 9.20 & 0.270 & 0.022 \\
& & 5 & 3.38 & 24.60 & 1.49 & 9.12 & 0.302 & 0.038\\ 
AudioTurbo \cite{zhao2025audioturbo}$^{**}$ & 1100M & 5 & - & 22.18 & 1.29 & 9.40 & 0.298 & - \\
\hdashline
\rowcolor{blue!10}
\multirow{1}{*}{MeanAudio (Ours)$^\diamondsuit$} & \multirow{1}{*}{120M} & \textit{1} & \textit{\textbf{1.77}} & \textit{\textbf{14.30}} & \textit{\textbf{1.32}} & \textit{\textbf{10.02}} & \textit{\textbf{0.290}} & \textit{\textbf{0.013}} \\
\rowcolor{blue!10}
& & 2 & 1.84 & 13.57 & 1.27 & 10.70 & 0.305 & 0.015 \\
\rowcolor{blue!10}
& & 5 & 2.12 & 13.43 & 1.25 & 11.18 & 0.314 & 0.024 \\
\rowcolor{blue!10}
& & 25 & 2.30 & \underline{13.38} &  \underline{1.25} &  11.23 & \textbf{0.319} & \textbf{0.083} \\
\rowcolor{blue!10}

\bottomrule
\end{tabular}
}
\caption{Objective evaluation results on AudioCaps test set. The single-step generation results are in \textit{italic}. The best multi-step and single-step generation results are \textbf{bolded}, the second-best are \underline{underlined}. 
$\ast$: The parameter count refers to the diffusion backbone, excluding text encoders, VAE, and vocoders.
$\dag$: The real-time factor (RTF) is evaluated over 100 generations samples on an NVIDIA RTX 3090.
$**$: Closed-source models, results are transcribed from the original paper. 
$\ddag$: Open-source models, results are re-evaluated using the official checkpoint. 
$\diamondsuit$: These models integrated CFG into training, resulting in \textit{True} NFEs, for other models, NFE should actually be doubled. 
\vspace{-0.3cm}
} 
\label{tab:main_results}
\end{table*}

\section{Experiments}
\subsection{Datasets}
We collect a large-scale audio-text dataset to train MeanAudio, including AudioCaps \cite{kim2019audiocaps}, WavCaps \cite{mei2024wavcaps}, AudioSet \cite{gemmeke2017audio}, VGGSound \cite{chen2020vggsound}, and LP-MusicCaps (MC \& MTT parts) \cite{doh2023lp}. 
Among them, only AudioCaps provides high-quality human-annotated textual captions, while others are weakly labeled using Large Language Models \cite{schulman2022introducing} or Multimodal Large Language Models \cite{chu2024qwen2}.
During training, we truncate all audio into 10 seconds. For audios longer than 10s, we crop at most 5 non-overlapping 10s segments. 
In total, we collect approximately 2.8M audio-text pairs for training, with a total duration of 8k hours. 
A detailed dataset description is given in Appendix~\ref{app:datasets}. 

We use the test split of AudioCaps as the evaluation set, which contains 957 audio clips. Each audio of the test set is paired with 5 textual captions, and we randomly select 1 caption for audio generation. 
To further evaluate MeanAudio's music generation capability, we test its performance on the test set of MusicCaps \cite{agostinelli2023musiclm}, with results reported in Appendix~\ref{app:musiccaps_eval}. 

\subsection{Implementation Details}

MeanAudio has $N_1=4$ multi-modal blocks and $N_2=8$ single-modal blocks. 
The hidden dimension of the transformer is set to 448, and the network has a total of 120M parameters. 
For integrated classifier-free guidance, We set $\omega=0.3$ and $\kappa=0.9$, resulting in an effective guidance scale of $\frac{\omega}{1-\kappa}=3$. 
We sample timesteps $(t,r)$ according to a logit-normal distribution \cite{esser2024scaling}, with $\mu=0.4$ and $\sigma=1$. 
Given a sampled pair, we assign the larger value to $t$ and the smaller value to $r$. Additionally, 75\% of the samples are randomly set with $r=t$ to mix the vanilla instantaneous flow field with the mean flow field. 

Following the curriculum described in Section~\ref{sec:training_strategy}, the model is first pre-trained on the full dataset for 400k iterations with the standard flow matching objective, and subsequently fine-tuned on AudioCaps for 200k iterations using the mixed flow objective.
During both stages, we use a learning rate of 1e-4 with a linear warm-up of 1,000 steps. A step decay schedule reduces the learning rate to 10\% of its original value at 80\% and 90\% of total steps.
Batch sizes are set to 256 and 32 for pre-training and fine-tuning, respectively.
All experiments are conducted on four NVIDIA RTX 3090 GPUs, requiring approximately 48 hours for pre-training and 22 hours for fine-tuning.

\begin{table}[t]
    \centering
    \resizebox{\linewidth}{!}{
    \begin{tabular}{lccll}
    \toprule
    \rowcolor{white}
        \textbf{Models} & \textbf{Size} & \textbf{NFE}  & \textbf{OVL↑} & \textbf{REL↑} \\
        \midrule
        \rowcolor{blue!10}
        MeanAudio (Ours)  & 120M & 1 & \textbf{4.03}\scriptsize{$\pm$0.72} & \textbf{4.29}\scriptsize{$\pm$0.71} \\
        ConsistencyTTA & 559M & 1 & 3.32\scriptsize{$\pm$0.95} & 3.41\scriptsize{$\pm$1.03} \\
        AudioLCM & 160M & 1 & 3.47\scriptsize{$\pm$0.89} & 3.31\scriptsize{$\pm$1.13} \\
        \midrule
        \rowcolor{blue!10}
        MeanAudio (Ours)  & 120M & 25 & \textbf{4.19}\scriptsize{$\pm$0.70} &  \textbf{4.44}\scriptsize{$\pm$0.66} \\
        TangoFlux & 516M & 50 & 4.02\scriptsize{$\pm$0.87} & 4.22\scriptsize{$\pm$0.85}  \\
        AudioLDM-L-Full & 739M & 200  & 3.25\scriptsize{$\pm$1.00} & 3.33\scriptsize{$\pm$1.08}\\
        GenAU-Large-Full & 1250M & 200 & 3.66\scriptsize{$\pm$0.90} & 3.79\scriptsize{$\pm$0.94} \\
    \bottomrule
    \end{tabular}
    }
    \vspace{-0.2cm}
    \caption{Subjective evaluation results}
    \label{tab:subjective_evaluation}
    \vspace{-0.5cm}
\end{table}

For subjective evaluation, we adopt standard TTA evaluation metrics: Fréchet Distance (FD), Fréchet Audio Distance (FAD), Kullback–Leibler Divergence (KL), Inception Score (IS), and CLAP score. 
For objective evaluation, we recruit ten audio professionals to carry out a rating process, following \cite{liu2023audioldm, liu2024audioldm2}. 
Specifically, the generated samples are rated based on overall quality (OVL) and relevance to the input text (REL) on a scale of 1 to 5.
To further evaluate the model's inference speed, we also report the Real-time Factor (RTF) of the system, which denotes the ratio between the total time a system takes to synthesize an audio and the duration of the audio. 
More details about evaluation can be found in Appendix \ref{app:evaluation}.

\subsection{Main Results}
We compare the performance of MeanAudio with other TTA models under both few-step ($< 10$ NFEs)  and multi-step ($\geq 10$ NFEs) generation. 
For few-step generation, we compare with SOTA accelerated TTA models. 
For multi-step generation, we compare with best-performing diffusion and flow-based models. 

As shown in Table \ref{tab:main_results}, for objective metrics, MeanAudio demonstrates SOTA performance under single-step generation, achieving an FD of 14.30, a KL of 1.32, an IS of 10.02, and a CLAP score of 0.290, outperforming all accelerated models such as ConsistencyTTA and AudioLCM by large margins.
Remarkably, its single-step generation performance even surpasses some multi-step baselines that require hundreds of function evaluations.
Moreover, MeanAudio achieves the fastest real-time factor of 0.013, representing a 100x speedup over the SOTA open-sourced diffusion-based model, GenAU, which has an RTF of 1.612.

As the number of sampling steps increases, MeanAudio’s generation quality also improves, revealing a trade-off between inference speed and output fidelity. Its two-step and five-step generations consistently outperform previous SOTA methods, highlighting its strong few-step synthesis capabilities. 
For multi-step generation, MeanAudio remains competitive with the best-performing audio systems. 
With only 25 synthesis steps, it achieves an FD of 13.38, a KL of 1.25, and a CLAP score of 0.318. 
It is worth noting that MeanAudio contains only 120M parameters, whereas other TTA systems typically exceed 500M.

Furthermore, the subjective evaluation results in Table \ref{tab:subjective_evaluation} align well with these objective findings.
For single-step generation, MeanAudio achieves the highest human-rated overall quality and relevance, obtaining OVL and REL scores of 4.03 and 4.29, respectively, surpassing ConsistencyTTA and AudioLCM.
Under the multi-step settings, MeanAudio attains an OVL of 4.19 and a REL of 4.44, outperforming all diffusion-based competitors. 
These evaluation results confirm that MeanAudio is capable of generating high-quality and semantically coherent audio, while being both fast and lightweight.

\subsection{Ablation Studies}
We conduct a comprehensive ablation study covering training strategies, architectural designs, and flow configurations to identify best practices for building a MeanFlow-based audio generator.
Here, all experiments except the training curriculum are trained from scratch on AudioCaps for 200k steps.

\begin{table}[t]
    \centering
    \resizebox{\linewidth}{!}{
    \begin{tabular}{lccccc}
    \toprule
        \textbf{Models} & \textbf{NFE} &  \textbf{FD↓} & \textbf{KL↓} &   \textbf{IS↑}& \textbf{CLAP↑} \\
        \midrule
        FluxAudio (Ours) & & 99.10 & 4.48 & 2.45 & -0.007 \\
        MeanAudio-Scratch & 1 & 16.13 & 1.36 & 9.62 & 0.285 \\
        \rowcolor{blue!10}
        \quad w. Pre-training &  & \textbf{14.30} & \textbf{1.32} & \textbf{10.02} & \textbf{0.290} \\
        
        \midrule
        FluxAudio (Ours) &  & 15.70 & 1.30 & \textbf{11.24} & \textbf{0.328} \\
         MeanAudio-Scratch & 25 & 14.41 & 1.32 & 10.79 & 0.314 \\ 
        \rowcolor{blue!10}
        \quad w. Pre-training & & \textbf{13.38} & \textbf{1.25} & 11.23 & 0.318 \\
    \bottomrule
    \end{tabular}
    }
    \vspace{-0.2cm}
    \caption{Ablation study of the training curriculum. 
    \vspace{-0.2cm}
    }
    \label{tab:ablation-curriculum}
\end{table}

\begin{table}[t]
    \centering
    \resizebox{\linewidth}{!}{
    \begin{tabular}{cccccc}
    \toprule
    \textbf{Ratio of $r=t$} 
        &  \textbf{NFE} &   \textbf{FD↓} &  \textbf{KL↓} &   \textbf{IS↑} & \textbf{CLAP↑}
        \\
        \midrule
        0\%  & \multirow{4}{*}{1}   & 138.34 & 5.91 & 1.19 & -0.061 \\
        25\% &                       & 23.73 & 1.75 & 7.30 & 0.213 \\
        50\% &                       & 16.18 & 1.39 & 9.45 & \textbf{0.287} \\
        \rowcolor{blue!10}
        75\% &                       & \textbf{16.13} & \textbf{1.36} & \textbf{9.62} & 0.285 \\
        \midrule
        0\%  & \multirow{4}{*}{25}  & 150.75 & 6.28 & 1.14 & -0.050 \\
        25\% &                       & 22.15  & 1.63 & 7.80 & 0.241 \\
        50\% &                       & 15.42  & 1.37 & 9.78 & 0.296 \\
        \rowcolor{blue!10}
        75\% &                       & \textbf{14.41} & \textbf{1.32} & \textbf{10.79} & \textbf{0.314} \\
    \bottomrule
    \end{tabular}
    }
    \caption{Ablation study of the flow mix-up ratio.}
    \vspace{-0.3cm}
    \label{tab:ablation-flowratio}
\end{table}

\noindent\textbf{Instantaneous-to-Mean Curriculum. }
We begin by studying the benefits of the proposed training strategy. 
For this, we first evaluate the performance of the model trained with the standard flow matching objective, denoted as FluxAudio. 
As shown in Table~\ref{tab:ablation-curriculum}, FluxAudio delivers strong performance in multi-step generation but performs poorly under single-step scenarios. 
For instance, it achieves a CLAP score of 0.328 with 25 NFEs, while its one-step synthesis yields only -0.007. 
This large performance gap arises because the vanilla flow matching objective guides only toward the instantaneous velocity field and small displacements.
Secondly, we trained MeanAudio using the mixed flow objective as described in Section~\ref{sec:training_strategy}, but without initializing it from FluxAudio, which we denote as MeanAudio-Scratch.
As shown in Table~\ref{tab:ablation-curriculum}, training MeanAudio with a mixed flow field significantly enhances its single-step generation, as the FD and IS improve to 16.13 and 9.62 at NFE = 1. 
Finally, training MeanAudio on top of FluxAudio with mixed flow fields further improves its performance in both single- and multi-step generation. 
As illustrated in Table~\ref{tab:ablation-curriculum}, at NFE = 1, the FD, CLAP score, and IS increase to 14.30, 10.02, and 0.290, respectively. Additionally, at NFE = 25, these scores rise to 13.38, 11.23, and 0.318, which are comparable to FluxAudio.  
These results demonstrate that our proposed instantaneous-to-mean curriculum can effectively enhance generation quality under both single- and multi-step scenarios.

\noindent\textbf{Flow Field Mix-up. }
We then investigated the effectiveness of the flow field mix-up by changing the ratio of $r=t$. Remember that this equals the percentage of using the vanilla flow matching objective during the fine-tuning stage.
As illustrated in Table \ref{tab:ablation-flowratio}, when the flow ratio $r=t$ is set to 0 (using only the mean flow objective), training becomes highly unstable and fails to converge. Setting the ratio to 25\% stabilizes training, but we observe only a modest improvement in multi-step generation performance, with FD increasing by just 2.5\%. 
As the flow ratio further increases, convergence accelerates, and generation performance also improves in both single- and multi-step settings. 
Specifically, when the ratio increases from 25\% to 75\%, the single-step CLAP score increases from 0.213 to 0.285, and the multi-step IS improves from 7.80 to 10.79. 
These results suggest that in audio latent modeling, jointly learning the instantaneous and mean velocity can stabilize training and enhance performance in both single-step and multi-step generation, as the two flows complement each other.

\noindent\textbf{Architectural Designs. }
We then study the effectiveness of different model components. 
As shown in Table~\ref{tab:ablation-architecture}, removing the CLAP encoder causes a clear drop in text–audio alignment, as the CLAP score drops from 0.285 to 0.270 and KL degrades from 1.36 to 1.42. These results underscore the importance of the dual-encoder design for semantic coherence. 
Similarly, excluding RoPE also reduces the CLAP score from 0.285 to 0.276, indicating that rotary positional embeddings can help preserve relative temporal order and strengthen alignment. 
Finally, removing ConvMLP degrades both the generation fidelity and diversity, as the FD score drops from 16.13 to 17.22 and IS worsens from 9.62 to 9.21, which highlights its role in temporal modeling and local feature extraction.
We also include a subjective evaluation of the dual-encoder design and experiments on the scalability of MeanAudio, which can be found in Appendix~\ref{app:exp_additional}.

\begin{table}[t]
    \centering
    \resizebox{\linewidth}{!}{
    \begin{tabular}{lccccc}
    \toprule
        \textbf{Models} & \textbf{NFE} &  \textbf{FD↓} & \textbf{KL↓} &   \textbf{IS↑}& \textbf{CLAP↑} \\
        \midrule
        \rowcolor{blue!10}
        MeanAudio (Ours) & \multirow{4}{*}{1} & 16.13 & \textbf{1.36} & 9.62 & \textbf{0.285} \\
        \quad w/o CLAP & & 16.21 & 1.42 & 9.80 & 0.270 \\
        \quad w/o RoPE & & \textbf{15.87} & 1.41 & \textbf{9.83} & 0.276 \\
        \quad w/o ConvMLP & & 17.22 & 1.38 & 9.21 & 0.283 \\
    \bottomrule
    \end{tabular}
    }
    \vspace{-0.2cm}
    \caption{Ablation study of the network design.}
    \vspace{-0.2cm}
    \label{tab:ablation-architecture}
\end{table}

\begin{table}[t]
    \centering
    \resizebox{\linewidth}{!}{
    \begin{tabular}{lccccc}
    \toprule
        \textbf{CFG Scale} & \textbf{NFE} &  \textbf{FD↓} & \textbf{KL↓} &   \textbf{IS↑}& \textbf{CLAP↑} \\
        \midrule
        1 (No CFG) &  & 28.02 & 2.03 & 4.90 & 0.185 \\
        2          &                    & 16.33 & 1.44 & 8.65 & 0.270 \\
        \rowcolor{blue!10}
        3          &          1         & \textbf{16.13} & \textbf{1.36} & 9.62 & \textbf{0.285} \\
        4          &                    & 16.23 & \textbf{1.36} & 9.52 & 0.284 \\
        5 & & 17.09 & 1.40 & \textbf{9.78} & 0.284 \\ 
    \bottomrule
    \end{tabular}
    }
    \caption{Ablation study of the CFG scale. 
    \vspace{-0.4cm}
    }
    \label{tab:ablation-cfgscale}
\end{table}

\noindent\textbf{Integrated Classifier-Free Guidance. }
We finally conducted an ablation study across different guidance scales to evaluate the effectiveness of CFG. As shown in Table~\ref{tab:ablation-cfgscale}, increasing the CFG scale from 1 (no guidance) to 3 leads to substantial improvements in both generation quality and semantic alignment, where FD, IS, and CLAP score have increased by 42.4\%, 96.3\%, and 54.1\%, respectively.
Further increasing the guidance scale to 4 and 5 yields marginal improvements in IS, but slightly degrades FD, KL, and CLAP, suggesting a trade-off between generation diversity and prompt adherence. 
These results suggest that a moderate guidance scale around 3 is the optimal configuration for single-step audio generation.
Note that the CFG is integrated into MeanAudio's training and incurs no additional cost during sampling.

\section{Conclusion}
In this work, we present MeanAudio, a novel MeanFlow-based fast and faithful text-to-audio generator. 
Built upon an enhanced Flux-style latent transformer, MeanAudio regresses the guided average velocity field during training, enabling fast generation by directly mapping from the start to the endpoint of the flow trajectory.
To enhance training stability and generation quality, MeanAudio adopts an instantaneous-to-mean curriculum with flow-field mix-up, facilitating stable optimization and rapid convergence.
Extensive experiments demonstrate that MeanAudio delivers SOTA performance in single-step generation and competitive results in multi-step synthesis. 
Comprehensive ablation studies further highlight the importance of architectural design, training curriculum, and flow configurations, providing valuable insights for developing faster and stronger audio systems.

\section*{Limitations}
While MeanAudio achieves strong performance in both single-step and multi-step generation, several limitations remain.
Firstly, the training data used for MeanAudio are cropped to 10-second segments, which may limit its ability to generate variable-length or long-form audio. 
Secondly, since MeanAudio is trained exclusively on public datasets, which can be noisy to some extent, its generation quality may still lag behind models trained on larger and cleaner proprietary datasets.
Thirdly, MeanAudio currently focuses on text-to-sound and music generation, and therefore lacks the ability to produce intelligible speech or handle fine-grained linguistic content.



\bibliography{custom}

\newpage
\appendix

\section{Related Work}
\subsection{Text to Audio Generation}
Text-to-Audio Generation (TTA) focuses on generating sound based on text inputs. Current TTA models are often based on the Latent Diffusion Model (LDM) \cite{rombach2022high} architecture. Among all, AudioLDM \cite{liu2023audioldm} pioneered this by training a U-Net to perform denoising conditioned on CLAP \cite{wu2023large} embeddings. Tango \cite{ghosal2023text} further improves the instruction following abilities by using the Large Language Model (LLM) FLAN-T5 \cite{chung2024scaling} as the text encoder. More recently, Stable-Audio-Open \cite{evans2025stable} adopted the Diffusion Transformer (DiT) \cite{peebles2023scalable} to generate variable-length, full-band audio samples. Meanwhile, LAFMA \cite{guan2024lafma} and TangoFlux \cite{hung2024tangoflux} integrated Flow Matching \cite{liu2022flow} to improve the efficiency and fidelity of audio generation.  
Additionally, recent work has explored Reinforcement Learning (RL) \cite{liao2024baton, majumder2024tango2, wang2025t2a} and Retrieval-Augmented Generation (RAG) \cite{yuan2024retrieval, yang2024audiobox} to enhance generation controllability and output quality.
While these models achieve high fidelity and controllability, they often suffer from slow inference due to the iterative sampling process inherent to diffusion- and flow-based models.

\subsection{TTA with Inference Acceleration}
To accelerate TTA inference, ConsistencyTTA first integrates Consistency Distillation \cite{song2023consistency} on TTA Latent Diffusion Models, where a teacher model, Tango \cite{majumder2024tango2}, provides supervision to train a distilled few-step student generator.
Subsequently, AudioLCM \cite{liu2024audiolcm} enriches this framework by employing a multi-step Ordinary Differential Equation (ODE) solver, while SoundCTM \cite{saito2025soundctm} introduces a novel feature distance to enable flexible single-step and multi-step generation. 
In parallel, Presto \cite{novack2024presto} proposes a dual-faceted distillation strategy that reduces both the number of sampling steps and model parameters to improve inference efficiency. FlashAudio \cite{liu2024flashaudio} and AudioTurbo \cite{zhao2025audioturbo} explored the use of Rectified Flow \cite{liu2022flow} and Rectified Diffusion \cite{wang2024rectified} to learn straight generative paths for rapid audio synthesis. Meanwhile, Stable-Audio-Small \cite{novack2025fast} employs contrastive post-training with adversarial loss to construct a compact few-step generator.
In this work, we investigate the use of MeanFlow to develop a TTA model that achieves strong performance in both single-step and multi-step generation.

Despite encouraging results, these distillation-based approaches often require significant computational resources, as online methods require holding 2-3 full models in memory at the same time, and offline methods rely on large-scale generation and storage of teacher trajectories before training.
Moreover, their performance is inevitably limited by the pre-trained teacher model and the consistency constraint do not provide properties of the underlying ground-truth field that should guide learning, which could lead to unstable training \cite{geng2025mean}.

\section{Evaluation Details}
\label{app:evaluation}
In this section, we first provide a detailed explanation of the objective evaluation metrics, followed by a description of our human evaluation process.

Among all metrics, FD and FAD measure the distance between the generated audio distribution and the real audio distribution. A low FD indicates that the generated audio is realistic and closely resembles the reference audio. KL evaluates how semantically similar the generated audio is to the reference audio. IS measures the diversity and quality of the generated samples, and CLAP score\footnote{The CLAP score is calculated based on the checkpoint:  \url{https://huggingface.co/lukewys/laion_clap/blob/main/music_speech_audioset_epoch_15_esc_89.98.pt}} measures how the generated audio align with the textual prompt. 
FD, IS, and KL are calculated based on the state-of-the-art audio tagger PANNs \cite{kong2020panns}, while FAD is calculated by VGGish \cite{hershey2017cnn}. 

\begin{figure}
    \centering
    \includegraphics[width=\linewidth]{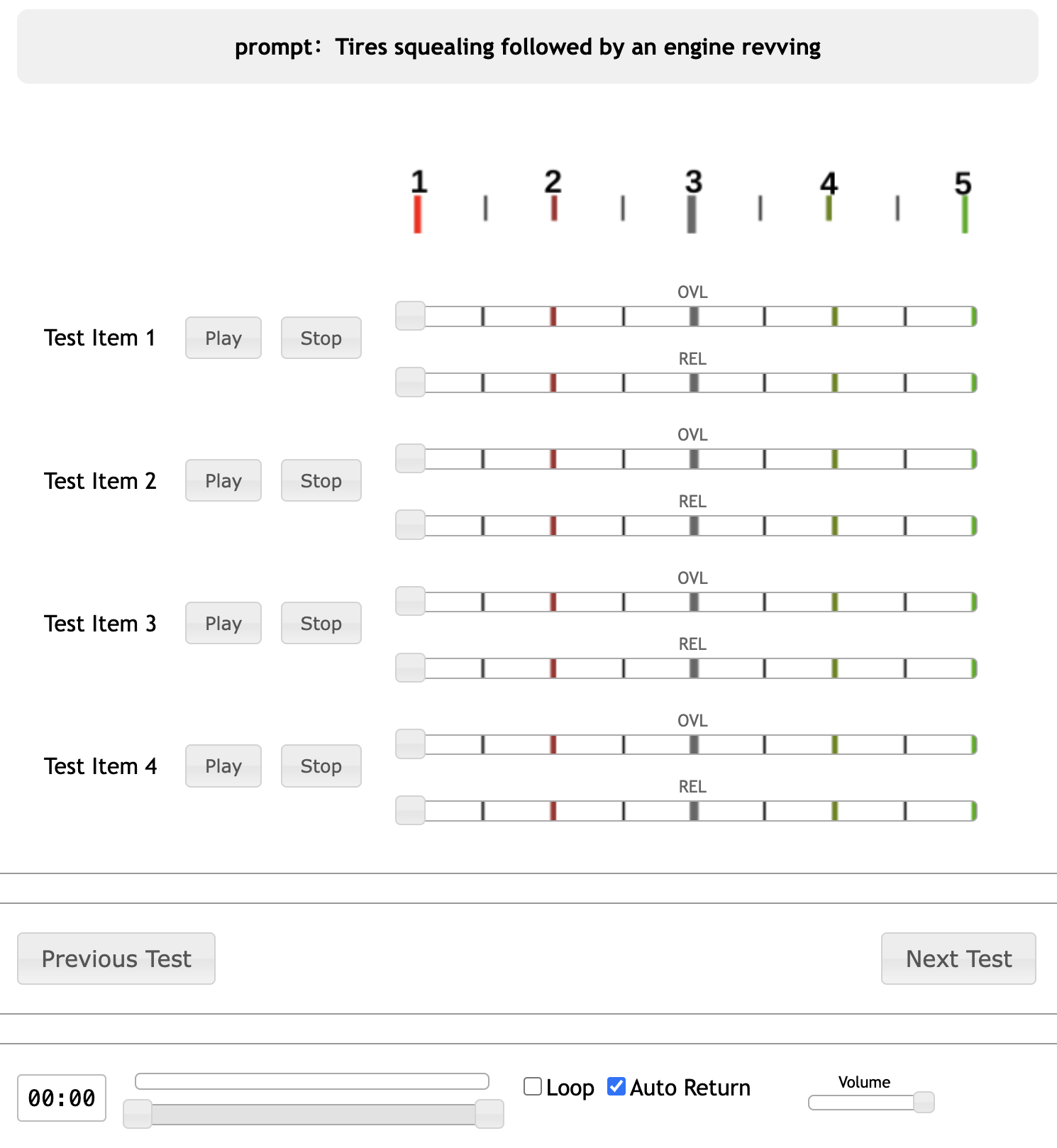}
    \caption{Screenshot of the subjective evaluation platform.}
    \label{fig:eval_plateform}
    \vspace{-0.5cm}
\end{figure}

For human subjective evaluation, we selected 10 audio experts to perform the listening test. 
Each participant is presented with 10 audio samples generated by different models along with their corresponding prompts. The captions are randomly selected from the AudioCaps test set, and the order of the audio samples is also randomized. Participants are asked to provide OVL and REL scores for each sample. An illustration of the evaluation platform is shown in Figure~\ref{fig:eval_plateform}. 

.

\begin{table*}[t]
\centering
\begin{tabular}{lccccccc}
\toprule
\textbf{Model} & \textbf{Prams} & \textbf{NFE} & \textbf{FAD} ↓ & \textbf{FD} ↓ &  \textbf{KL} ↓ & \textbf{IS} ↑ & \textbf{CLAP}  ↑\\
\midrule

MeanAudio-S-AC & 120M & & \textbf{1.77} & 14.30 & 1.32 & 10.02 & 0.290 \\
MeanAudio-S-Full & 120M & 1 & 2.85 & \textbf{14.19} & 1.27 & 10.69 & 0.314 \\
MeanAudio-L-Full  & 480M &  & 2.17 & 14.22 & \textbf{1.23} & \textbf{11.24} & \textbf{0.316} \\
\midrule 
MeanAudio-S-AC & 120M & & \textbf{2.30} & \textbf{13.38} &  1.25 &  11.23 & 0.319 \\
MeanAudio-S-Full & 120M & 25 & 3.51 & \textbf{13.38} & 1.22 & 12.11 & 0.334 \\
MeanAudio-L-Full & 480M & & 2.59 & 14.06 & \textbf{1.21} & \textbf{12.46} & \textbf{0.339} \\
\rowcolor{blue!10}

\bottomrule
\end{tabular}
\caption{Experiments on Data \& Model Scaling. Results are evaluated on AudioCaps test set. 
}
\label{tab:exps_scaling}
\end{table*}

\section{Addtional Experiments}
In this section, we present several additional experiments, including a detailed human evaluation of the dual-encoder architecture, an analysis of the scalability of MeanAudio, and an assessment of its music generation capability.

\label{app:exp_additional}
\subsection{Human Evaluation on Dual Encoders}
\label{app:exps_encoder}
To more comprehensively evaluate and understand the benefits of the dual-encoder architecture, we designed a more detailed human evaluation protocol.
We first constructed a refined evaluation subset, aiming to better reflect the model’s ability to follow long and complex textual descriptions.
Specifically, we selected captions from the AudioCaps test set that contain more than 20 words, forming a subset named AudioCaps-Long.
We observed that such captions typically describe at least four distinct sound events, posing a greater challenge for models in terms of instruction comprehension and multi-event generation.
From AudioCaps-Long, we then randomly sampled 10 captions and asked 3 human listeners to evaluate the relevance (REL) of the generated audio outputs.
During this evaluation, the listeners were instructed to pay particular attention to fine-grained sound events and their temporal order when assigning scores.

\begin{table}[t]
    \centering
    \resizebox{\linewidth}{!}{
    \begin{tabular}{lcccl}
    \toprule
    \rowcolor{white}
        \textbf{Models} & \textbf{Size} & \textbf{NFE}  & \textbf{Encoder} & \textbf{REL↑} \\
        \midrule
        MeanAudio  & \multirow{1}{*}{120M} & \multirow{1}{*}{1} & T5+CLAP & \textbf{4.10}\scriptsize{$\pm$0.37} \\
        MeanAudio  & 120M & 1 & T5 & 3.75\scriptsize{$\pm$0.51} \\
    \bottomrule
    \end{tabular}
    }
    \vspace{-0.2cm}
    \caption{Subjective evaluation on dual encoders}
    \label{tab:sub_evaluation_encoder}
    \vspace{-0.5cm}
\end{table}

As shown in Table~\ref{tab:sub_evaluation_encoder}, using both FLAN-T5 and CLAP consistently yields better performance than employing T5 alone.
The dual-encoder model tends to capture subtle sound events more accurately, such as a dog barking at the beginning or a child crying toward the end of the audio, confirming the effectiveness of our architectural design.

\begin{table*}[t]
\centering
\resizebox{1\linewidth}{!}{
\begin{tabular}{lccccccc}
\toprule
\textbf{Model} & \textbf{Prams}$^\ast$ & \textbf{NFE} & \textbf{FAD} ↓ & \textbf{FD} ↓ &  \textbf{KL} ↓ & \textbf{IS} ↑ & \textbf{CLAP}  ↑ \\
\midrule
\textbf{Accelerated TTA Models} \\
AudioLCM \cite{liu2024audiolcm}$^\dag$  & 160M & 1 & 5.98 & 40.68 & 1.62 & 2.33 & 0.149 \\
ConsistencyTTA \cite{bai2023consistencytta}$^{\dag}$  & 559M & 1 & 4.89 & 67.30 & 1.87 &1.69 & 0.138  \\
\rowcolor{blue!10}
\multirow{1}{*}{MeanAudio-S-AC (Ours)}$^1$ & 120M  & 1 & 3.38 & 27.96 & 1.71 & 2.29 & 0.200 \\
\rowcolor{blue!10} 
\multirow{1}{*}{MeanAudio-S-Full (Ours)}$^2$ & 120M & 1 & \textbf{1.26} & \textbf{11.82} & \textbf{1.11} & \textbf{2.90} & \textbf{0.302} \\

\midrule
\textbf{Diffusion- and Flow-based TTA Models} 
\\
FluxMusic \cite{fei2024flux}$^\ddag$  & 2100M & 200 & 1.43 & - & 1.25 & 2.98 & - \\
AudioLDM-2-Large \cite{liu2024audioldm2}$^\ddag$    & 712M  & 200   & 2.93 & 16.34  & 1.40 & 2.59   & -\\ 
Stable-Audio-Open \cite{evans2025stable}$^\ddag$  & 1050M & 200 &  3.51 & 36.42 & 1.56 & 2.93 & - \\
ETTA \cite{lee2024etta}$^\ddag$  & 1440M & 100& 1.91 & 10.06 & \textbf{1.04} & \textbf{3.32} & - \\
\rowcolor{blue!10}
MeanAudio-S-AC (Ours) & 120M & 25 & 3.05 & 24.73 & 1.71 & 2.56 & 0.221 \\
\rowcolor{blue!10}
MeanAudio-S-Full (Ours) & 120M & 25 & \textbf{1.21} & \textbf{10.05} & 1.08 & 3.01 & \textbf{0.315} \\

\bottomrule
\end{tabular}
}
\caption{Objective evaluation results on MusicCaps test set. $\dag$: Results are evaluated using the officially released checkpoint. $\ddag$: Results are transcribed from the original paper. $^1$: MeanAudio-S-AC: Model initialized from FluxAudio and fine-tuned with the mixed flow objective on the training set of AudioCaps. $^2$: Model initialized from FluxAudio and fine-tuned with the mixed flow objective on the full training set. Detailed explanation about these models found in Appendix~\ref{app:exps_scale}. 
\vspace{-0.3cm}
} 
\label{tab:results_mc}
\end{table*}

\subsection{Experiments on Scalability}
\label{app:exps_scale}
We further investigate the scalability of MeanAudio with respect to both data and model size.
For data scaling, we initialize MeanAudio from FluxAudio and train it on the entire 2.8M-sample dataset using the mixed flow objective for 400k iterations with a batch size of 256.
The resulting large-scale variant is denoted as MeanAudio-S-Full.
For model scaling, we increase the hidden dimension to 896, resulting in a model with 448M parameters, referred to as MeanAudio-L-Full.
We compare these two new variants with MeanAudio-S-AC, which was initialized from FluxAudio and was trained on AudioCaps with the mixed-flow objective. 

As shown in Table~\ref{tab:exps_scaling}, both data scaling and model scaling substantially improve MeanAudio’s generation quality.
When trained on the full 2.8M-sample dataset, the model (MeanAudio-S-Full) achieves clear gains over MeanAudio-S-AC under both single-step and multi-step settings.
At NFE = 1, the IS and CLAP scores increase from 10.02 to 10.69 and 0.290 to 0.314, respectively, indicating stronger prompt adherence and improved synthesis fidelity.
Further scaling the model size to 480M parameters (MeanAudio-L-Full) yields consistent improvements, with IS = 11.24 and CLAP = 0.316, surpassing all other single-step audio generators and approaching the performance of multi-step systems.
A similar trend is observed in the multi-step scenario, where scaling enhances both fidelity and semantic alignment: MeanAudio-L-Full achieves CLAP = 0.339 and IS = 12.46, the best among all configurations.
These results demonstrate that MeanAudio can benefit from both data and model scaling, confirming its scalability and capacity to leverage larger datasets and architectures for higher-quality audio generation.

\subsection{Text-to-Music Generation}
\label{app:musiccaps_eval}
Finally, we evaluate MeanAudio's music generation abilities, where we assess its performance on the test split of MusicCaps \cite{agostinelli2023musiclm}. As summarized in Table \ref{tab:results_mc}, MeanAudio significantly outperforms existing accelerated text-to-audio systems while achieving performance on par with or superior to larger diffusion- and flow-based models.

Under single-step generation, MeanAudio-S-AC achieves an FD of 27.96 and an IS of 2.29, significantly surpassing AudioLCM and ConsistencyTTA by large margins.
By further scaling the training data, MeanAudio continues to exhibit substantial improvements in music generation performance. Specifically, MeanAudio-S-Full achieves an FD of 11.82 and an IS of 2.90, showcasing its enhanced capability to synthesize high-quality music within a single generation step.

When the sampling steps are increased to 25, MeanAudio-S-Full further improves to a FAD of 1.21, FD of 10.05, and CLAP of 0.315, reaching performance comparable to state-of-the-art diffusion-based music generation systems that typically require hundreds of iterations.
These results highlight the strong generative capability of MeanAudio and demonstrate the effectiveness of the MeanFlow framework in producing perceptually rich music signals.

\begin{table*}
    \centering
    \begin{tabular}{ccccc}
    \rowcolor{white}
    \toprule \textbf{Dataset Name} & \textbf{\# Clips } & \textbf{Caption Source} & \textbf{Duration} \\
    \midrule 
      AudioCaps \cite{kim2019audiocaps}  & 49k & Human \cite{kim2019audiocaps} & 130h \\
      AudioSet \cite{gemmeke2017audio}   & 1.7M & MLLM$^\dag$ \cite{bai2025audiosetcaps} & 4722h  \\
      FreeSound &  612k &  LLM$^\ddag$ \cite{mei2024wavcaps} & 1700h \\
    BBC-Sound-Effects & 121k & LLM \cite{mei2024wavcaps} & 336h  \\
    AudioSet-Strong \cite{hershey2021benefit} & 106k & LLM \cite{mei2024wavcaps} & 294h \\
      VGGSound \cite{chen2020vggsound} & 178k & Human \cite{chen2020vggsound} & 494h \\
      LP-MusicCaps-MC & 2.3k & LLM \cite{doh2023lp} & 6.3h \\
      LP-MusicCaps-MTT & 43k & LLM \cite{doh2023lp} & 119h \\
      \midrule
      Total & 2.8M & - & 7801h \\
    \bottomrule
    \end{tabular}
    \caption{Dataset Statistics. $^\dag$ MLLM: Multimodal Large Language Models. $^\ddag$ LLM: Large Language Models. } 
    \label{tab:datases_statistics}
\end{table*}

\section{Dataset Details}
\label{app:datasets}
In this section, we present a detailed description of the datasets used in the training of MeanAudio. Table \ref{tab:datases_statistics} provides an overall statistics of all datasets. 

\noindent\textbf{AudioCaps. }AudioCaps \cite{kim2019audiocaps} contains approximately 50k audio-text pairs, where each audio contains one human-labelled high-quality caption.

\noindent\textbf{WavCaps. }
WavCaps \cite{mei2024wavcaps} comprises 400k audio samples collected from multiple sources, including BBC Sound Effects,
\footnote{\url{https://sound-effects.bbcrewind.co.uk}}, 
FreeSound
\footnote{\url{https://freesound.org}}
, 
SoundBible
\footnote{\url{https://soundbible.com}} 
and AudioSet-Strong \cite{hershey2021benefit}.
Each audio has one caption generated by ChatGPT. 

\noindent \textbf{AudioSet. }
AudioSet \cite{gemmeke2017audio} contains approximately 2M audio samples. However, these audio segments only contain ground-truth labels. We use the caption from AudioSetCaps \cite{bai2025audiosetcaps} to train our TTA models. 

\noindent \textbf{VGGSound. }
VGGSound \cite{chen2020vggsound} contains about 200k audio samples. Like AudioSet, the audio is only paired with labels. We directly use labels to train the TTA models. 

\noindent \textbf{LP-MusicCaps. }
LP-MusicCaps \cite{doh2023lp} is a large-scale weakly-labeled music caption dataset, sourced from MusicCaps (MC) \cite{agostinelli2023musiclm}, MagnaTagATune (MTT) \cite{law2009evaluation}, and Million-Song-Dataset (MSD). 
We use the MC and MTT splits of LP-MusicCaps, containing a total of 25k music segments. 

\end{document}